\documentclass[aps,prx,floatfix,longbibliography,twocolumn,superscriptaddress]{revtex4-2}

\usepackage{graphicx}
\usepackage{enumerate}
\usepackage{subfigure}
\usepackage{dsfont}
\usepackage{xcolor}
\usepackage{qcircuit}
\usepackage{hyperref}
\hypersetup{
  colorlinks=true,
  citecolor=blue
}
\usepackage{amssymb}
\usepackage{amsmath}
\usepackage{mathtools}
\usepackage{amsthm}
\DeclarePairedDelimiter\bra{\langle}{\rvert}
\DeclarePairedDelimiter\ket{\lvert}{\rangle}
\newcommand{\Tr}{\mathrm{Tr}}
\newtheorem{theorem}{Theorem}
\newtheorem{definition}{Definition}
\begin{document}

\title{Predicting Gibbs-State Expectation Values with Pure Thermal Shadows}

\author{Luuk Coopmans}
\email{luuk.coopmans@quantinuum.com}
\affiliation{Quantinuum, Partnership House, Carlisle Place, London SW1P 1BX, United Kingdom}

\author{Yuta Kikuchi}
\affiliation{Quantinuum K.K., Otemachi Financial City Grand Cube 3F, 1-9-2 Otemachi, Chiyoda-ku, Tokyo, Japan}

\author{Marcello Benedetti}
\affiliation{Quantinuum, Partnership House, Carlisle Place, London SW1P 1BX, United Kingdom}

\date{June 26, 2023}

\begin{abstract}
The preparation and computation of many properties of quantum Gibbs states is essential for algorithms such as quantum semidefinite programming and quantum Boltzmann machines. We propose a quantum algorithm that can predict $M$ linear functions of an arbitrary Gibbs state with only $\mathcal{O}(\log{M})$ experimental measurements. Our main insight is that for sufficiently large systems we do not need to prepare the $n$-qubit mixed Gibbs state explicitly but, instead, we can evolve a random $n$-qubit pure state in imaginary time. The result then follows by constructing classical shadows of these random pure states. We propose a quantum circuit that implements this algorithm by using quantum signal processing for the imaginary time evolution. We numerically verify the efficiency of the algorithm by simulating the circuit for a ten-spin-1/2 XXZ-Heisenberg model. In addition, we show that the algorithm can be successfully employed as a subroutine for training an eight-qubit fully connected quantum Boltzmann machine.  
\end{abstract}

\maketitle

\section{Introduction}
Gibbs states are mixed quantum states that describe quantum systems in thermodynamic equilibrium with their environment at finite temperature. They play a central role in quantum statistical mechanics~\cite{Gogolin2016} and their properties are of importance for a wide range of academic and industry-relevant applications~\cite{Alhambra2022}; these include the design of complex quantum materials in condensed-matter physics, optimization with quantum semidefinite programming, and machine learning with quantum Boltzmann machines.

The preparation of Gibbs states and the computation of Gibbs-state expectation values are highly nontrivial tasks. Existing algorithms can have rather complicated implementations and may apply only to a limited set of systems. Classical algorithms suffer from the exponentially growing Hilbert space or have a sign problem for some fermionic systems~\cite{Loh1990}. Fault-tolerant quantum algorithms have better asymptotic scaling~\cite{Yung_2012, Chowdhury_2016} but have qubit and noise requirements that exceed the capabilities of current and near-term quantum processors.
Variational quantum algorithms can prepare Gibbs states~\cite{Endo_2020, Liu_2021} and cope with some hardware limitations but require many experimental measurements for each optimization step and may suffer from barren plateaus~\cite{McClean2018}. 
Other quantum approaches based on minimally entangled typical thermal states (METTS)~\cite{White2009, Motta2020} set up a Markov chain that potentially has a long thermalization time. 

Gibbs states can be represented mathematically by a density matrix of the form $\rho_\beta=e^{-\beta H}/Z$, where $H$ is the system Hamiltonian, $\beta$ is the inverse temperature, and $Z=\Tr{ e^{-\beta H} }$ is the partition function. In this work, we propose an efficient quantum algorithm to estimate a large number of Gibbs-state expectation values without preparing and measuring the mixed state $\rho_\beta$. This is achieved by combining thermal \emph{pure} quantum (TPQ) states~\cite{Bocchieri1959,Hams2000,Iitaka2003,Popescu_2006,Sugiura2012,Sugiura2013,Hyuga2014,Endo2018,Jin_2021,Iwaki_2021,Hongo2021} and classical shadow tomography~\cite{Ohliger_2013,Aaronson2017,Paini2019,Huang2020,Levy2021,Kunjummen2021,McGinley2022,Elben2022}. One way to generate a thermal pure quantum state is by imaginary time evolution, $e^{-\beta H/2}\ket{\phi}$, of a $n$-qubit random state $\ket{\phi}$~\cite{Richter2021,Powers2021}. We show that if the probability distribution over the initial states, $\ket{\phi}$, forms at least a quantum 2-design, the imaginary-time-evolved states approximate the expectation values of $\rho_\beta$ up to an error that falls off exponentially with the system size, $n$.

\begin{figure}
\subfigure{
    \includegraphics[width=0.94\columnwidth]{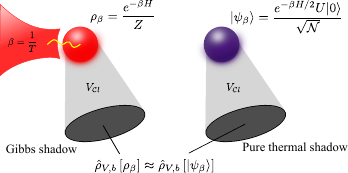}
}
\subfigure{
    \hspace{0.5cm}
    \Qcircuit @C=1em @R=1.em {
        \lstick{\text{QSP: } \ket{000}} & {/}\qw & \qw & \multigate{1}{\mathcal{U}_\text{QSP}} & \rstick{\ket{000}}\qw 
        \\
        \lstick{\text{System: } \ket{0}^{\otimes n}} & {/}\qw & \gate{U_{\mathcal{C}l}}&\ghost{\mathcal{U}_\text{QSP}}& \gate{V_{\mathcal{C}l}} &\meter{} & \hspace{1.2cm} \Rightarrow\text{Shadow}
    }
}
\caption{Top: The classical shadows of a Gibbs state $\rho_\beta$, here represented as a quantum system (red sphere) in thermal equilibrium with its environment (red bath), are equal to the classical shadows constructed from a thermal pure quantum state $\ket{\psi_\beta}$ (purple sphere). The random measurement direction of the shadows is determined by the Clifford unitary $V_{\mathcal{C}l}$.  Bottom: A circuit diagram of the quantum circuit that implements a pure thermal shadow with quantum signal processing (QSP).} 
\label{fig:1}
\end{figure}

Classical shadow tomography is then exploited to construct an efficient classical representation of these TPQ states from outcomes of randomized measurements. Using the results from Ref.~\cite{Huang2020}, we show that for sufficiently large systems, we can estimate $M$ Gibbs-state expectation values using $\mathcal{O}(\log{M})$ measurements of a single prepared TPQ state. This is remarkable, because the required number of measurements is similar to the case where we have actually prepared the Gibbs state (e.g., via a purification~\cite{Chowdhury_2016}, which incurs a higher cost of $2n$ qubits).

We also propose a practical implementation of the algorithm. We begin by preparing the initial state using a polynomial-depth Clifford circuit. This suffices to produce a quantum 2-design with the additional benefit of a reduced circuit depth compared to some of the other random-state preparation techniques. We then approximate the imaginary time evolution using quantum signal processing (QSP)~\cite{Low_2016,Low_2017,Low_2019, Martyn_2021_grand}. This approach is very general (i.e.,  it applies to any quantum Hamiltonian $H$, in principle) andit offers great control since we can systematically trade off circuit depth for accuracy. For the randomized measurement, we use another Clifford circuit. The complete circuit, comprising a Clifford circuit, a QSP circuit, and a randomized measurement, is shown in Fig.~\ref{fig:1} and uses a total number of qubits linear in the system size, $n$. 

In order to support our theoretical findings, we classically simulate the quantum circuit for a quantum system of ten spin-1/2 particles in the XXZ-Heisenberg model~\cite{Heisenberg1928, Franchini_2017}. Moreover, we use the simulated algorithm to numerically train a fully connected quantum Boltzmann machine for the generative modeling of both classical and quantum data to high accuracy.     

\section{Thermal Pure Quantum States}
A TPQ state is any pure state, $\ket{\psi}$, that is able to estimate a fixed set of properties (expectation values) of a sufficiently large mixed thermal state, specified by some specific statistical ensemble. For the thermodynamic (canonical) Gibbs ensemble, following Ref.~\cite{Sugiura2013}, we define it to be any pure state $\ket{\psi}$, which is drawn at random, that satisfies 
\begin{equation}
\label{eq:purethermalstate}
    \mathrm{Pr}\left[|\bra{\psi}O_j\ket{\psi} - \Tr{\rho_\beta O_j}|\geq \epsilon\right]\leq C_\epsilon e^{-\alpha n},
\end{equation} 
for all $O_j$ in some predefined set of Hermitian operators~$\{O_j\}$. The constants $C_\epsilon$ and $\alpha$ are specified later.

We now show that the pure states generated by imaginary time evolution, 
\begin{equation}
\label{eq:cliff_ite_state}
    \ket{\psi_\beta} = \frac{e^{-\beta H/2}U\ket{0}}{\sqrt{\bra{0}U^\dagger e^{-\beta H}U\ket{0}}} \equiv \frac{e^{-\beta H/2}U\ket{0}}{\sqrt{\mathcal{N}}},
\end{equation} 
where $U\in \mathcal{C}l(2^n)$ is a random unitary drawn from the $n$-qubit Clifford group, satisfy Eq.~(\ref{eq:purethermalstate}).
These TPQ states are slightly different from the ones originally proposed in Ref.~\cite{Sugiura2013} and also differ from the ones used in Ref.~\cite{Jin_2021} for classical numerical simulations. There, a particular form of Haar-random states for $U\ket{0}$ is used, which on a quantum computer would require exponential circuit depth. Our choice of $U\in \mathcal{C}l(2^n)$ yields a unitary 3-design~\cite{Webb2015} and thus it replicates Haar integrals up to the third moment. This has the advantage of requiring only $\mathcal{O}(n^2/\log{n})$ quantum gates~\cite{Aaronson2004}.

First, as shown in detail in Appendix~\ref{app:TPQ}, the expectation value of an arbitrary Hermitian operator $O$ in the random pure states $\ket{\psi_\beta}$ is, on average,
\begin{equation}
    \mathbb{E}_U[\bra{\psi_\beta}O\ket{\psi_\beta}] \approx \Tr{ \rho_\beta O } + \Tr{ \rho_\beta^2 } ( \Tr{\rho_\beta O} - \Tr{\rho_{2\beta} O}).
\label{eq:etherm}
\end{equation}
Here, $\mathbb{E}_U$ denotes the ensemble average with respect to the $n$-qubit Clifford group $U\in\mathcal{C}l(2^n)$. The approximate equality follows from computing a Taylor expansion up to second order in the variance of the normalization $\mathcal{N}$ of $\ket{\psi_\beta}$ (for details, see Appendix \ref{app:TPQ}). This leads to the same approximation as in previous works based on Haar-random states~\cite{Sugiura2013, Jin_2021}.

A similar calculation for the variance of the expectation value with respect to $U$ yields
\begin{equation}
\begin{split}
    &\text{Var}[\bra{\psi_\beta}O\ket{\psi_\beta}] \approx  \\ 
    &\quad \Tr{\rho_\beta^2} \Big( 
    \frac{\Tr{(Oe^{-\beta H})^2}}{\Tr{e^{-2 \beta H}}} -
     2 \Tr{\rho_\beta O}\Tr{\rho_{2\beta} O} + (\Tr{\rho_\beta O})^2 \Big).
\label{eq:var}
\end{split}
\end{equation}

Both the bias in Eq.~(\ref{eq:etherm}) and the variance in Eq.~(\ref{eq:var}) are proportional to the purity of the Gibbs state, $\Tr{\rho_\beta^2}$. Since the Gibbs state $\rho_\beta$ minimizes the Helmholtz free energy, $F_\beta = - \log{Z}/\beta$, we can write
\begin{equation}
\label{eq:purity}
    \Tr{\rho_\beta^2} = \frac{\Tr{e^{-2\beta H}}}{Z^2} 
    = e^{-2\beta(F_{2\beta}-F_\beta)}
    =\mathcal{O}(e^{-n}). 
\end{equation}
The last equality follows from properties of the free energy, i.e., the extensivity, $F_\beta\propto n$,~\footnote{We assume that the system size is large enough for the contributions from boundaries to be negligible.} and the monotonicity with respect to the inverse temperature, $F_{2\beta} > F_\beta$. The terms that multiply the purity in Eqs.~(\ref{eq:etherm}) and (\ref{eq:var}) have the form of expectation values and can be bounded by the spectral norm, $\|O\|^2$. This means that for operators with polynomially large $\|O\|^2$, the variance vanishes exponentially with $n$~\footnote{This includes most observables that one typically considers in a physical setting, e.g., any $k$-local Pauli operator.}. After application of Markov's inequality (see Appendix~\ref{app:TPQ}) we find that $\ket{\psi}=\ket{\psi_\beta}$ satisfies Eq.~(\ref{eq:purethermalstate}) with $C_\epsilon \approx 4\|O\|^2/\epsilon^2$ and $\alpha n = 2\beta(F_{2\beta}-F_\beta)$.

Hence, the expectation values with respect to the random pure states $\ket{\psi_\beta}$ can be used as estimators for Gibbs-state expectation values of polynomially sized operators $O$, for a sufficiently large system and finite $\beta$. Importantly, for $\beta=\infty$, i.e.,  when the Gibbs state approaches the ground state of $H$, the Gibbs state becomes pure, $\Tr{\rho_\beta^2}=1$, and the error remains finite for any system size $n$. For all other $\beta$, the rate of exponential decay, and thus how large $n$ needs to be, is dictated by $F_{2\beta}-F_\beta$. The exact error, therefore, depends on which specific system Hamiltonian $H$ one considers, the inverse temperature $\beta$, and also the spectral norm of the observable $\|O\|$. This means that in some specific instances~\cite{Sugiura2013, Jin_2021},the use of only a single TPQ state is sufficient (see Appendix~\ref{app:add_num}). In contrast, approaches that do not have a vanishing variance, such as algorithms based on METTS~\cite{White2009, Motta2020}, may require multiple pure states. 

\section{Pure Thermal Shadows}
Classical shadow tomography~\cite{Ohliger_2013,Aaronson2017,Paini2019,Huang2020} is a method to efficiently predict many properties of a prepared quantum state, $\rho$, by performing randomized measurements and constructing efficient classical representations (shadows) from the measurement outcomes. In Ref.~\cite{Huang2020}, it has been proven that one can predict $M$ properties of $\rho$ from $\mathcal{O}(\log{M})$ measurements (for a review, see Appendix~\ref{app:shadows}). For this reason, classical shadow tomography is particularly beneficial in scenarios where one needs to compute many observables, such as in variational quantum algorithms and quantum fidelity estimation~\cite{Elben2022}.

To predict many properties of a Gibbs state $\rho_\beta$, one can prepare a purification of $\rho_\beta$ on $2n$ qubits and apply classical shadow tomography. Here, instead, we simplify the process by constructing shadows of the $n$-qubit TPQ states $\ket{\psi_\beta}$. We refer to these as \emph{pure thermal shadows}. Notably, in the limit of large system size, the number of state preparations and measurements required for predicting $M$ properties to a certain accuracy becomes similar to the case where we prepare and measure $\rho_\beta$ directly.

The first step of the algorithm is the application of a random unitary, $V$, to the TPQ states, $\ket{\psi_\beta}\mapsto V\ket{\psi_\beta}$. The operation $V$ can either be a random $n$-qubit Clifford unitary, $V\sim\mathcal{C}l(2^n)$, for a random \emph{Clifford} measurement, or a tensor product of single-qubit Clifford unitaries, $V\sim\mathcal{C}l(2)^{\otimes n}$, for a random \emph{Pauli} measurement.
Subsequently, a computational-basis measurement of $V\ket{\psi_\beta}$ is performed to obtain a $n$-qubit bit-string outcome, $\ket{b}$. The thermal shadows,
\begin{equation}
    \hat{\eta}_{V,b} = \mathcal{M}^{-1}\left(V^\dagger \ket{b}\bra{b}V\right),
\end{equation} 
can be constructed with $\mathcal{M}^{-1}(X) = (2^n+1)X - \mathds{1}\Tr{X}$. As $V$ is a Clifford circuit and $\ket{b}$ a bit string, the shadow can be efficiently constructed and stored classically.

By averaging over both the unitary $V$ of the randomized measurements and the unitary $U$ of the TPQ states $\ket{\psi_\beta}$, we find that the expectation values of the pure thermal shadows satisfy
\begin{equation}
    \mathbb{E}_{U}\mathbb{E}_{V,b}\left[ \Tr{\hat{\eta}_{V,b} O}\right] = \mathbb{E}_{U}\left[\Tr{\ket{\psi_\beta}\bra{\psi_\beta}O}\right].
\end{equation}
Here, $\mathbb{E}_{V,b}$ includes both the classical average over the random circuits $V$ as well as the quantum average of the measurement outcomes $\ket{b}$ via Born's probabilities. The expectation value of the shadow $\hat{\eta}_{V,b}$ is thus the same as the expectation value of $\ket{\psi_\beta}$. By Eqs.~(\ref{eq:etherm})~and~\eqref{eq:purity}, this becomes equal to the Gibbs-state expectation value, $\Tr{\rho_\beta O}$, as we increase $n$. 

The success probability of the algorithm depends on the mean squared error, $\sigma^2$, of the pure thermal-shadow expectation value from the true Gibbs-state expectation value. For random Clifford measurements, $V\sim\mathcal{C}l(2^n)$, we find (see Appendix~\ref{app:proof}) that this error evaluates to
\begin{align}
\label{eq:shadowvar}
\sigma^2 &= \mathbb{E}_{U}\mathbb{E}_{V,b}[(\Tr{\hat{\eta}_{V,b}O} -  \Tr{\rho_\beta O} )^2]
    \nonumber\\
    &\approx \Tr{O_0^2} +2\Tr{\rho_\beta O_0^2} - \left(\Tr{\rho_\beta O_0}\right)^2 + \mathcal{O}(e^{-n}).
\end{align}
Here, we define $O_0=O-\mathds{1}\Tr{O}/2^n$ as the traceless part of the operator $O$.
Similarly, for random Pauli measurements, $V\sim \mathcal{C}l(2)^{\otimes n}$, we obtain
\begin{align}
\label{eq:shadowvar_pauli}
\sigma^2
    \approx 3^k - (\Tr{\rho_\beta O})^2 + \mathcal{O}(e^{-n}),
\end{align}
where $O$ is now a $k$-local Pauli operator~\footnote{For a general $k$-local operator $O$, the Pauli mean squared error is bounded as $\sigma^2 \le 4^k\lVert O_0\rVert^2 - (\mathbb{E}_{U}[\bra{\psi_\beta}O_0\ket{\psi_\beta}])^2 + \mathcal{O}(e^{-n}) $.}

Remarkably, these mean squared errors are, up to an exponentially small bias, approximately the same as the $\sigma^2$ of shadows constructed from randomized Pauli and Clifford measurements of a prepared $\rho_\beta$ (see Appendix~\ref{app:proof}). This means that the construction of shadows of the TPQ state performs approximately as well as the construction of shadows of the true Gibbs state. However, the latter scenario is only theoretical, since up to now no quantum device has been able to prepare a desired Gibbs state exactly.

In the scenario in which we wish to predict a total of $M$ Gibbs-state expectation values, we invoke a specific form of Theorem $1$ in Ref.~\cite{Huang2020} for a median-of-means estimator. This gives the main result of this paper, which can be summarized by the following theorem.

\begin{theorem}
\label{thm:1}
Given $M$ Hermitian operators $O_1, \dots, O_M$, and accuracy parameters $\epsilon, \delta \in[0,1]$. A total of
\begin{equation}
   n_{\mathrm{shadows}} = NK = \frac{27\log(M/\delta)}{\epsilon^2} \max_j \left[\sigma_j^2\right]
\end{equation}
classical shadows $\hat{\eta}_{V, b}$ of a thermal pure quantum state $\ket{\psi_\beta}$, with mean squared error $\sigma_j^2$ [Eq.~\eqref{eq:shadowvar} or~\eqref{eq:shadowvar_pauli}] for observable $O_j$, is sufficient to estimate all $M$ Gibbs-state expectation values, $\Tr{\rho_\beta O_j}$, with a success probability of at least $1-\delta$, i.e.,
\begin{equation}
    \mathrm{Pr}\Big[\max_j|\mu_j(N,K) - \Tr{\rho_\beta O_j}|\leq \epsilon\Big]\ge 1-\delta.
\end{equation}
\end{theorem}

Here, $\mu_j(N,K)$ is the predicted expectation value for observable $O_j$ obtained from a median-of-means estimation on $K$ subsets of $N$ classical shadows. A detailed proof of this Theorem and a review of Theorem 1 in Ref.~\cite{Huang2020} are given in Appendices~\ref{app:shadows} and~\ref{app:proof}.
We conclude that for $M$ Gibbs-state expectation values, we require $\mathcal{O}(\log{M})$ preparations of TPQ states $\ket{\psi_\beta}$ and randomized measurements. The exact performance of the algorithm depends on which type of measurement one uses (Pauli or Clifford) and the considered observables. For $k-$local Pauli observables, with $k\ll n$, the variance of Pauli measurements [Eq.~\eqref{eq:shadowvar_pauli}], is lower and hence a better choice. On the other hand, for tasks such as fidelity estimation, the Clifford measurements perform better~\cite{Huang2020}. 

\section{Implementing the Quantum Circuit}
In order to implement the pure thermal-shadow algorithm on a quantum device we need several different elements. First, we need a method to sample uniformly from the $n$-qubit Clifford group to generate the random Clifford circuits $U$. An efficient polynomial-time algorithm exists for this~\cite{Bravyi_2021}, which has been implemented in the quantum computing library \textsc{qiskit}. We can use this for both the generation of the random pure states at the beginning of the algorithm as well as the randomized measurements at the end. Second, we need a method for approximating the nonunitary operator $e^{-\beta H /2}$ in Eq.~\eqref{eq:cliff_ite_state}. To this end, we use quantum signal processing~\cite{Low_2019} (QSP), a framework for performing matrix arithmetics on quantum computers.

QSP can generate polynomials of a Hamiltonian given access to a block encoding of the Hamiltonian into a larger unitary matrix (see Appendix~\ref{app:qsp}). There exist block-encoding schemes for generic matrices~\cite{Low_2017,Gilyen_2019}. Here, we employ the linear combination of unitaries (LCU) method~\cite{Childs2012}, under the assumption that the Hamiltonian is given in the form $H = \sum_k a_k P_k$, where $a_k \in \mathbb{R}$ and $P_k$ are $n$-qubit Pauli operators. We also require a preprocessing step that rescales the spectrum of $H$ to the interval $[0,1]$. We achieve this with a min-max rescaling $\tilde{H} = (H - \lambda_\text{min} \mathds{1})/(\lambda_\text{max} - \lambda_\text{min})$, where $\lambda_\text{min}$ ($\lambda_\text{max}$) is the smallest (largest) eigenvalue.
In practice, the extremal eigenvalues are unknown and one resorts to a lower bound for $\lambda_\text{min}$ and to an upper bound for $\lambda_\text{max}$. We also define the imaginary time $\tau = \beta (\lambda_\text{max} - \lambda_\text{min} )/2$ so that
\begin{equation}
    \label{eq:ite_op}
    e^{- \tau \tilde{H} } = e^{\beta \lambda_\text{min} / 2} e^{- \beta H /2}  .
\end{equation}
Thus, we can evolve $\tilde{H}$ for time $\tau$ and obtain the desired nonunitary operator up to a constant factor. This factor is irrelevant, as it cancels out in Eq.~\eqref{eq:cliff_ite_state}.

We then use QSP to implement a polynomial approximation to the exponential function and apply it to the eigenvalues of $\tilde{H}$~\cite{Gilyen_2019,van_Apeldoorn_2020}.
The method in Corollary 64 of Ref.~\cite{Gilyen_2019} approximates the exponential function to accuracy $\epsilon$ with a polynomial of degree $\mathcal{O}(\sqrt{\beta(\lambda_\text{max} - \lambda_\text{min})} \log(1/\epsilon))$. The degree is proportional to the number of uses of the block-encoding circuit and thus determines the overall depth of the circuit. In our implementation, we instead numerically find a polynomial approximation to $e^{-\tau |x|}$ using the \textsc{pyqsp} library~\cite{Martyn_2021_grand} and then construct a suitable QSP circuit.

At the bottom of Fig.~\ref{fig:1}, we show the circuit diagram for the pure thermal-shadow algorithm.
The success probability $\| e^{- \tau \tilde{H}} U \ket{0} \|^2$, i.e., the probability of measuring the ancilla qubits in the all-zero state, depends on the random Clifford circuit $U$. Averaging over the Clifford group and using Eq.~\eqref{eq:ite_op}, we find that the probability of success is $\mathcal{O}(e^{\beta \lambda_\text{min}} \Tr e^{-\beta H}/2^n)$.
It can be verified that the success probability decreases as $\beta$ increases. This is expected from the intuition that low-temperature sampling is computationally harder~\cite{Bravyi2021b}.  By including $\mathcal{O}(e^{-\beta \lambda_\text{min}/2} \sqrt{2^n / \Tr e^{-\beta H}})$ iterations of amplitude amplification, the protocol is enhanced and succeeds with probability $\mathcal{O}(1)$. Amplitude amplification can be implemented via yet another layer of QSP~\cite{Martyn_2021_grand}. Note that while this protocol works for any Hamiltonian and temperature, the total circuit depth is expected to be polynomial in $n$ only for certain choices of $H$ and $\beta$.

\section{Numerical Simulations}
\subsection{Verification of the efficiency of the algorithm}
We verify the efficiency of the pure thermal-shadow algorithm by simulating the circuit for the one-dimensional XXZ-Heisenberg Hamiltonian~\cite{Heisenberg1928, Franchini_2017} given by 
\begin{equation}\label{eq:xxz}
    H =\sum_{i=1}^{n-1}J\left(\sigma^{x}_i\sigma^{x}_{i+1}+\sigma^y_i\sigma^y_{i+1}\right)+\Delta \sigma_i^z\sigma^z_{i+1}.
\end{equation} 
Here, $J$ is the nearest-neighbor $x,y$-coupling constant and $\Delta$ is the nearest-neighbor $z$-coupling strength. This is a paradigmatic model for the study of (quantum) magnetism and it maps to the Hubbard model for spinless interacting fermions after a Jordan-Wigner transformation. We set the system size to $n=10$ and the coupling parameters to $J=0.5$ and $\Delta=0.75$, which means that the ground state is in the antiferromagnetic phase~\cite{Franchini_2017}. 

\begin{figure}
    \includegraphics[width=0.95\columnwidth]{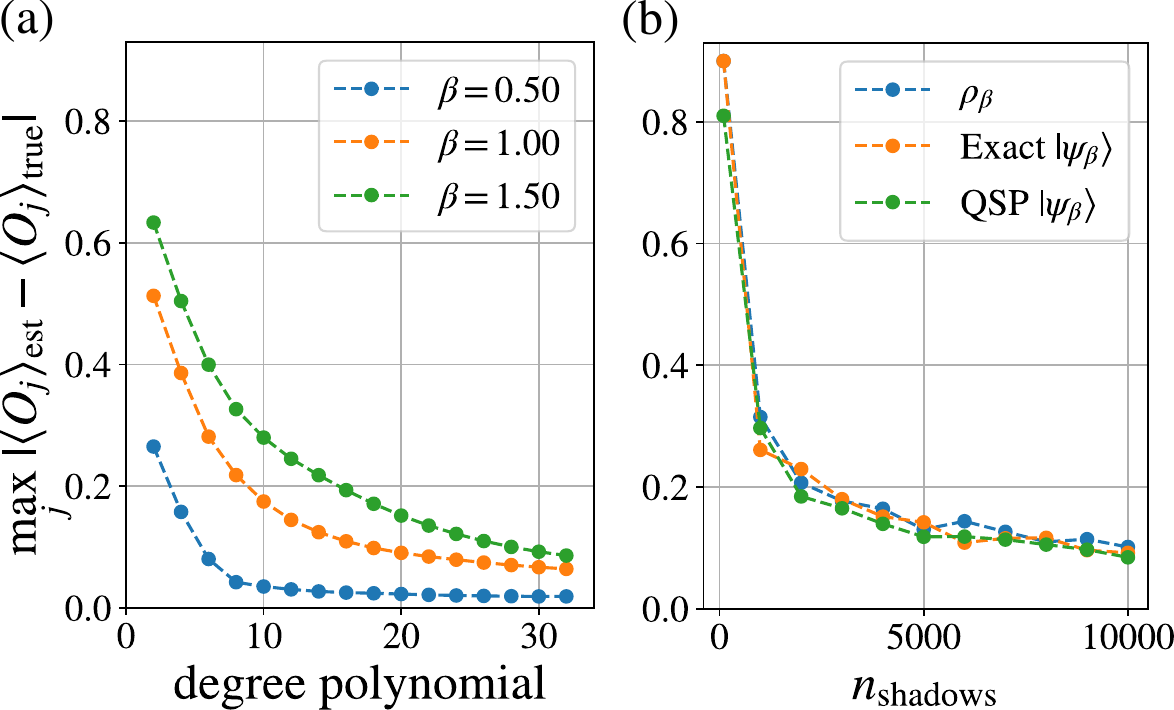}
    \caption{The maximum error between the exact Gibbs-state expectation values, $\langle O_j\rangle_{\mathrm{true}} = \Tr{O_j \rho_\beta}$, and the estimated expectation values $\langle O_j\rangle_{\mathrm{est}}$ for all possible one- and two-qubit Pauli operators $O_j$. (a) The expectation values estimated directly on thermal pure quantum states, $|\psi_\beta\rangle$, generated with QSP for different inverse temperatures $\beta$ as function of the degree of the polynomial approximation. (b) A comparison of the errors between shadows constructed directly from the true Gibbs state, $\rho_\beta$, and shadows constructed from the exact TPQ states and the TPQ states generated with QSP: $\beta=1$ and polynomial degree $d=32$.}
    \label{fig:shadow_numerics}
\end{figure}

First, in Fig.~\ref{fig:shadow_numerics}(a), we show the maximum error of the estimated expectation values of all one- and two-qubit Pauli observables (435 in total) using the TPQ states in Eq.~\eqref{eq:etherm} generated with QSP. Note that here we use the full TPQ state vectors to estimate the expectation values and do not construct thermal shadows yet. We observe that the error decreases as we increase the degree of the polynomial approximation used in QSP. This shows that we can use QSP to tune the accuracy. In addition we see that, as expected, the method performs better for higher temperatures, i.e.,  smaller $\beta$. 

In Fig.~\ref{fig:shadow_numerics}(b), we show the same error computed for the full algorithm, in which we use the mean expectation values of the thermal shadows to estimate the true Gibbs-state expectation values. As we are computing Pauli observables, we use random Pauli measurements for this. We compare the performance of shadows constructed from the exact Gibbs state $\rho_\beta$ with $\beta=1$, the exact TPQ state, and the TPQ state obtained from a QSP of degree 32. As we increase the number of shadows used to estimate the expectation values, we observe that the maximum error becomes similar, $\mathcal{O}(10^{-1})$, for all three methods. This confirms our theoretical finding that shadows constructed from $\ket{\psi_\beta}$ perform as well as shadows constructed from~$\rho_\beta$.

\subsection{Training of quantum Boltzmann machines}
In order to demonstrate the power of the pure thermal-shadow algorithm for practical applications, we exploit it for the training of a quantum Boltzmann machine (QBM). A QBM is a quantum machine-learning model, $\rho_\theta=e^{-H(\theta)}/Z$ (see Appendix~\ref{app:add_num}), that can be used for various learning tasks~\cite{Amin_2018, Benedetti2017, Kieferova2016, Kappen2020}, e.g., the generative modeling of classical and quantum data and Hamiltonian learning via Gibbs states~\cite{Anshu_2021}. The training of a QBM is, however, highly complex and intractable, since most optimization algorithms require the evaluation of many Gibbs-state expectation values, $\Tr{\rho_\theta O_j}$. Here, we use the pure thermal-shadow algorithm in this subroutine.

We focus on an eight-qubit fully connected QBM, where the model Hamiltonian has the form $H(\theta) = \sum_{k=x,y,z}\sum_{i,j>i}^{n}\lambda^k_{ij}\sigma_i^k\sigma^k_j + \sum_i^n \gamma^k_i\sigma^k_i,$
with $\theta\equiv(\lambda,\gamma)$ the model parameters. As objective function, we take the quantum relative entropy, $S(\eta \| \rho_\theta) = \text{Tr}(\eta \log{\eta}) - \text{Tr}(\eta \log{\rho_\theta})$, where $\eta$ is the target density matrix that encodes our data.

\begin{figure}
    \includegraphics[width=0.95\columnwidth]{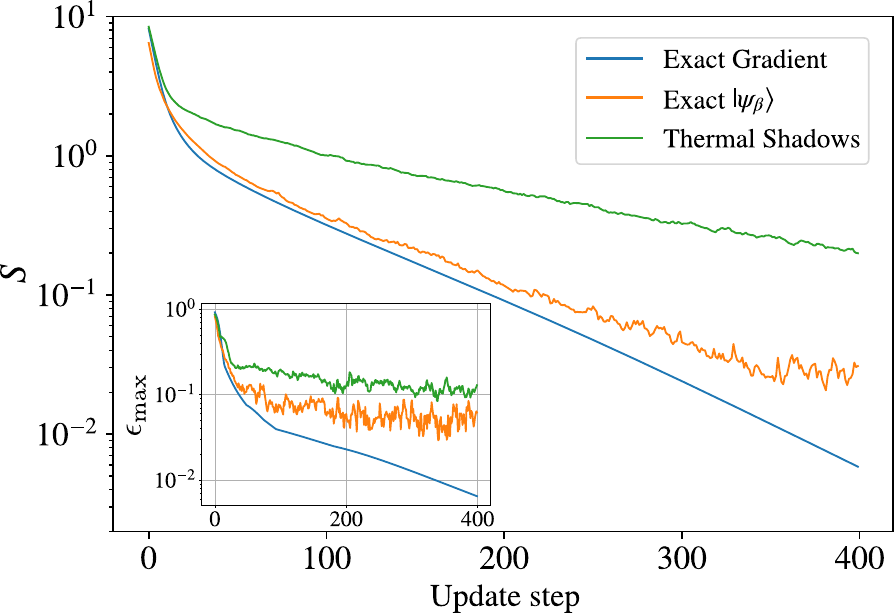}
    \caption{The quantum relative entropy, $S(\eta \| \rho_\beta)$, versus the gradient-descent update step. The three curves are for three different methods with which the required Gibbs-state expectation values in the gradient are computed. The inset shows the maximum error between the expectation values of the target density matrix and the QBM, $\epsilon_{\mathrm{max}}\equiv\mathrm{max}_j|\langle O_j\rangle_{\rho_\theta} - \langle O_j \rangle_{\eta}|$, as a function of the update step. For these simulations, we use 5000 thermal (Pauli) shadows, a polynomial degree of 32, and a learning rate of $0.1$.}
    \label{fig:xxz_qbm_results}
\end{figure}

First, focusing on quantum data, we set $\eta$ equal to the Gibbs state of the XXZ Hamiltonian from Eq.~\eqref{eq:xxz} at $\beta=1$ and minimize $S$ with respect to $\theta$ using vanilla gradient descent. In Fig.~\ref{fig:xxz_qbm_results}, we show results for three different training methods. When using the exact gradient, both the relative entropy (main panel) and the error in the expectation values (inset) decrease monotonically during training. When using either the exact TPQ state or thermal shadows, we observe a similar behavior but with slower rates. This is expected, since these methods provide approximations to the gradient only up to a certain accuracy. 

Importantly, the maximum error in the expectation values of the final QBM from the target $\eta$ for the thermal-shadow method is $\epsilon_{\mathrm{max}}=\mathcal{O}(10^{-1})$. The error can be further reduced by increasing the number of shadows and/or the degree of polynomial approximation used for QSP. This shows that thermal shadows can be used to train a QBM to model quantum data to high accuracy. Moreover, our theoretical analysis indicates that the accuracy may become higher for increased system sizes. This is because the exact TPQ states are expected to better approximate the true Gibbs state for larger systems. 

\begin{figure*}[t]\centering
    \includegraphics[width=1.0\textwidth]{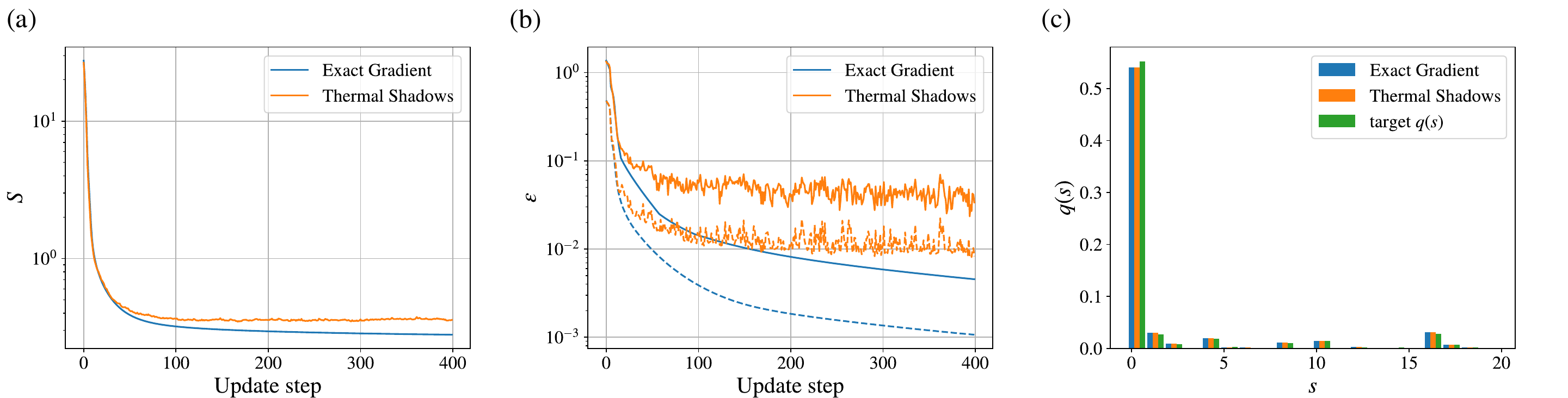}
    \caption{(a) The quantum relative entropy of the salamander retina target state with respect to the eight-qubit fully connected QBM during training. The training is performed with both the exact gradient and the pure thermal-shadows algorithm. For both methods, the relative entropy decreases monotonically until a small finite value is reached. For these simulations we use $10000$ thermal (Pauli) shadows, a QSP of degree $32$, and a learning rate $\alpha=0.2$. (b) The maximum error (solid lines) in the predicted expectation values, $\epsilon_{\mathrm{max}}\equiv\mathrm{max}_j|\langle O_j\rangle_{\rho_\theta} - \langle O_j \rangle_{\eta}|$, during training. For comparison, we also plot $\epsilon_{\mathrm{mean}}\equiv\mathrm{mean}|\langle O_j\rangle_{\rho_\theta} - \langle O_j \rangle_{\eta}|$, with dashed lines. These quantities also decrease and reach a finite value. The thermal-shadow method achieves a mean accuracy of order $\mathcal{O}(10^{-2})$. This accuracy is expected to improve if we increase the number of thermal shadows used during training. (c) The empirical probability distribution $q(s)$ of the target classical distribution obtained from the salamander ganglion cells data set (green) versus the distribution $\mathrm{diag}(\rho_\theta)$ obtained from the trained QBM. For visualization purposes, we only plot the probabilities of the first $20$ out of the total $2^8$ bit strings $s$. The distribution of the trained QBMs matches the target $q(s)$ closely.}
    \label{fig:qs}
\end{figure*}

As a second example, we train a fully connected QBM on a multielectrode array recording of eight salamander retinal-ganglion cells~\cite{Tkacik2014}~\footnote{This data set contains data for 160 ganglion cells, but we focus on the 8 cells that have the highest firing rate.}. In order to encode this classical data set into a quantum state $\eta$, we follow Ref.~\cite{Kappen2020}. First, we construct the empirical probability distribution, $q(s)=\frac{1}{N}\sum_{\mu=1}^N \delta_{s,s^\mu}$, where the $s$ are bit strings of length 8 and the $\{s^\mu\}_{\mu=1}^N$ are binary vectors of length 8 corresponding to the $N$ salamander data samples. This distribution is shown in Fig.~\ref{fig:qs}(c) for the first 20 bit strings. Then, we turn $q(s)$ into a rank-1 density matrix, $\eta=\ket{\psi}\bra{\psi}$, with $\ket{\psi}=\sum_s \sqrt{q(s)} \ket{s}$. The advantage of this encoding is that the required target expectation values can be efficiently computed classically and it imposes nontrivial quantum statistics in the classical data~\cite{Kappen2020}.

In Fig.~\ref{fig:qs}(a), we show the results for a trained QBM with this classical data encoding. We observe that the training algorithm based on pure thermal shadows performs almost as well as the one based on exact gradients. In contrast with the monotonically decreasing relative entropy observed for the quantum target state, the relative entropy now saturates at a small but finite value. This is expected, because the target $\eta$ is not a Gibbs state itself, i.e, there is a model mismatch. Note that this is not related to our training method, but a feature of the chosen model.

In Fig.~\ref{fig:qs}(b), we plot the maximum and mean error in the predicted expectation values, $\epsilon = |\langle O_j\rangle_\eta-\langle O_j\rangle_{\rho_\theta}|$, of the QBM during training. Similar to the XXZ-Heisenberg-model results, we see that during training the maximum error decreases and reaches a finite value for the thermal-shadow method. The final error is small, $\mathcal{O}(10^{-2})$, and can be further improved by increasing the number of shadows.

As a last indicator for the performance of our QBM training algorithm, in Fig.~\ref{fig:qs}(c), we inspect the final probability distribution of the QBM. It can be seen that the probability distribution obtained from the QBM approximates the true $q(s)$ to high accuracy. This shows that the algorithm can be exploited for the training of quantum machine-learning models on real-world classical data sets. We leave the question of how well a QBM performs compared to other generative machine-learning models for future studies.

\section{Conclusion and outlook}
In this work, we show that one can predict $M$ linear properties of a $n$-spin Gibbs state at finite temperature with only $\mathcal{O}(\log{M})$ preparations of a thermal pure quantum state. In practice, this means that one only needs to prepare a $n$-qubit random pure state on a quantum computer, imaginary time evolve it with the system Hamiltonian, and finally perform randomized measurements. Compared to algorithms that prepare a purification of the Gibbs state, our approach reduces the required qubits by half; and compared to variational Gibbs-state preparation, it reduces the number of experimental measurements.  

We also propose a practical implementation of the algorithm where the imaginary time evolution is performed by quantum signal processing. By simulating a XXZ-Heisenberg model we demonstrate the performance of the algorithm and circuit and show that the predicted expectation values are in excellent agreement with the exact values. In addition, as a relevant real-world application, we show that the algorithm can be exploited to train a fully connected quantum Boltzmann machine to model quantum and classical data with high accuracy. This problem is known to be computationally intractable.  

For future studies, a few interesting avenues can be explored. First, on a theoretical level, one can look at constructing shadows of other types of thermal pure states, such as states that satisfy the eigenstate-thermalization hypothesis~\cite{Srednicki1994, Gogolin2016} and minimally entangled typical thermal states~\cite{White2009}. Another option is to generalize the algorithm to other thermodynamic ensembles and look at the efficiency. 

More practically, one can investigate further improvements of the algorithm by derandomization~\cite{Hadfield2022, Huang_2021b,Zhang2021} of the measurements and also the prediction of nonlinear functions of $\rho_\beta$. Furthermore, it will be interesting to see if one can improve the imaginary-time-evolution approximation by quantum signal processing; e.g., by using fragmentation as described in Ref.~\cite{Silva2021}. We believe that this may lead to a practical realization of our algorithm, and the training of quantum Boltzmann machines, on early-term quantum signal processors. 

\section*{Acknowledgments} We thank David Amaro, Samuel Duffield, Masaru Hongo, Onno Huijgen, Oscar Watts, Bert Kappen, Michael Lubasch, Stefano Mangini, and Kirill Plekhanov for helpful discussions. We thank Mattia Fiorentini, Nathan Fitzpatrick, Matthias Rosenkranz, and Chris Self for providing feedback on an earlier version of this paper.

\appendix

\onecolumngrid

\renewcommand\thefigure{A\arabic{figure}}  
\renewcommand\thetable{A\arabic{table}}  
\setcounter{figure}{0}  
\setcounter{table}{0}
\setcounter{section}{0}

\section{Review of classical shadows}\label{app:shadows}

In this appendix we review the classical-shadow-tomography protocol as introduced by Huang \emph{et al.} in Ref.~\cite{Huang2020}.\footnote{Note that a related shadow-tomography protocol has been introduced by Aaronson~\cite{Aaronson2017}, which requires us to know the measurement operators prior to measurement. Furthermore, a form of randomized measurements has been discussed in Ohliger \emph{et al.}~\cite{Ohliger_2013} and a practical protocol in Paini and Kalev~\cite{Paini2019}.} This method uses randomized measurements to construct efficient classical representations (shadows) of an arbitrary quantum state $\rho$. 
The power of this randomized approach is that it can be used to predict many properties (such as observables) of $\rho$ with very few measurements. We start by outlining all the steps in this randomized measurement method (by following Ref.~\cite{Huang2020}) and defining what the classical shadows are. Afterward, we prove how many shadows are required to compute $M$ expectation values, $\Tr{\rho O_j}$, for a set of arbitrary observables $\{O_j\}_{j=1}^{M}$. 

\subsection{Randomized measurements and classical shadows}

\begin{definition} [Randomized measurements]
Given an arbitrary $n$-qubit quantum state $\rho$ and a random unitary operator $U$, which is drawn uniformly from some ensemble $\mathcal{U}$, we define a random measurement process as follows: 
\begin{equation} 
\rho\mapsto U\rho U^\dagger \mapsto |b\rangle\langle b|, 
\end{equation} 
where in the second step we measure in the computational basis, i.e.,  $b$ is a bit string. The probability of obtaining bit string $b$ is given by Born's rule: 
\begin{equation}\label{eq:Born}
\Tr[ U\rho U^\dagger |b\rangle\langle b|]
=\langle b|U\rho U^\dagger|b\rangle.
\end{equation}
\end{definition}
The interpretation of this measurement process is straightforward: we rotate the quantum state to some random basis and then perform a computational-basis measurement. Taking these two steps together, we thus perform a measurement in a random basis. 

This random measurement process can then be used to construct efficient classical representations of $\rho$ by performing some (classical) postprocessing steps. First, one applies the inverse of $U$ to the outcome $|b\rangle \langle b|$. This can be done efficiently classically if $U$ is drawn, for example, from the $n$-qubit Clifford group $\mathcal{C}l(2^n)$ or a tensor product of random single-qubit Cliffords, $U=U_1\otimes...\otimes U_n$, where $U_i\in \mathcal{C}l(2)$.
We then repeat this measurement process many times, each time for a different unitary $U$, and look at the expectation value, $\mathbb{E}[U^\dagger|b\rangle \langle b|U]$. We can see that this complete process defines a quantum channel $\mathcal{M}$ on the original state $\rho$. Specifically, we obtain
\begin{equation}
\label{eq:mchannel}
\mathbb{E}\left[U^\dagger|b\rangle \langle b|U\right] = \mathbb{E}_{U\in\mathcal{U}}\sum_{b\in \{0, 1\}^{n}}\langle b|U\rho U^\dagger|b\rangle U^\dagger|b\rangle \langle b|U\equiv \mathcal{M}(\rho).
\end{equation} 

As $\mathcal{M}$ is a linear map, we can obtain information about $\rho$ from the measurement results by inverting the channel $\mathcal{M}^{-1}$. We remark that the channel is invertible provided that the map is one to one. This means that there exist a unitary $U\in\mathcal{U}$ and a corresponding $b$ for which $\langle b|U\rho U^\dagger |b\rangle\neq\langle b|U\sigma U^\dagger |b\rangle$ for $\rho\neq\sigma$. The application of $\mathcal{M}^{-1}$ to both sides of Eq.~\eqref{eq:mchannel} gives 
\begin{equation}
    \mathbb{E}\left[\mathcal{M}^{-1}\left(U^\dagger|b\rangle \langle b|U\right)\right] = \rho. 
    \label{eq:recrho}
\end{equation} 
Note that here we use the fact that $\mathcal{M}^{-1}$ is a linear map and hence can be absorbed into the expectation value. From this, we can define the classical shadows.

\begin{definition}[Classical shadow]
Given a $n$-qubit quantum state $\rho$ and a bit string $\ket{b}$ obtained from a randomized measurement of $\rho$ with random unitary $U\in\mathcal{U}$, a classical shadow of the state $\rho$ is defined to be the unit-trace operator 
\begin{equation}
\label{eq:shadow}
    \hat{\rho}_{U,b} = \mathcal{M}^{-1}\left(U^\dagger|b\rangle \langle b|U\right),
\end{equation} 
provided that there exists a unitary $U\in\mathcal{U}$ and a corresponding $b$ for which $\langle b|U\rho U^\dagger |b\rangle\neq\langle b|U\sigma U^\dagger |b\rangle$ for $\rho\neq\sigma$.
\end{definition}

By definition, these classical shadows reproduce the quantum state $\rho$ if we average over the unitaries $U$ and measurement outcomes $b$, i.e., $\mathbb{E}[\hat{\rho}_{U,b}]=\rho$. However, in order to efficiently store and compute these shadows classically, we need to derive the explicit form of the channel $\mathcal{M}$. We do this in the next subsection for random Clifford unitaries $U\in\mathcal{C}l(2^n)$.  

\subsection{The inverse quantum channel\texorpdfstring{, $\mathcal{M}^{-1}$,}{} for randomized Clifford measurements}

To construct $\mathcal{M}^{-1}$, we evaluate the effect of the channel $\mathcal{M}$ on an arbitrary state $\rho$, 
\begin{align}
    \mathcal{M}(\rho) 
    =  \mathbb{E}_{U\in\mathcal{U}}\sum_{b\in \{0, 1\}^{n}}\langle b|U\rho U^\dagger|b\rangle U^\dagger|b\rangle \langle b|U 
    = \sum_{b\in \{0, 1\}^{n}}\mathbb{E}_{U\in\mathcal{U}}\langle b|U\rho U^\dagger|b\rangle U^\dagger|b\rangle \langle b|U.
\end{align} 
The evaluation of this expression for any arbitrary ensemble $\mathcal{U}$ is very hard, if possible at all. However, we recognize that the argument of the average over the ensemble $\mathcal{U}$ is a polynomial of order 2 in $(U, U^\dagger)$. This means that if the unitary $U$ is at least a unitary 2-design~\cite{Ambainis2007}, we can replace the inner average by an integral with respect to the Haar measure $d\mu_{\text{Haar}}(U)$,
\begin{equation}
    \sum_{b\in \{0, 1\}^{n}}\mathbb{E}_{U\in\mathcal{U}_{k\geq 2}}\langle b|U\rho U^\dagger|b\rangle U^\dagger|b\rangle \langle b|U =\sum_{b\in \{0, 1\}^{n}}\int_{U\in\mathcal{U}}d\mu_{\text{Haar}}(U) \langle b|U\rho U^\dagger|b\rangle U^\dagger|b\rangle \langle b|U ,
\end{equation}
where $\mathbb{E}_{U\in\mathcal{U}_{k\geq2}}$ is the average over a unitary $k$-design. An example of a category of unitary 3-designs, and thus also 2-designs, is the random Clifford unitaries, $U\in\mathcal{C}l(2^n)$~\cite{Webb2015}, which can be classically simulated and sampled efficiently by the Gottesmann-Knill theorem~\cite{Gottesman1997,Gottesman1998}. For this reason, from now on we use random $n$-qubit Clifford measurements. However, in reality we can use any randomized measurements with a unitary $k\geq2$-design.

The advantage of writing the average as a Haar integral is that expressions exist to evaluate some integrals analytically. 
In particular, below we repeatedly make use of the following three Haar-integral identities~\cite{Huang2020}:
\begin{align}
    \label{eq:Haar_1st}
    \int_{U\in\mathcal{U}}d\mu_{\mathrm{Haar}}(U)  \langle b|U O_1U^\dagger|b\rangle
    &= \frac{\Tr{O_1}}{2^n}. 
    \\
    \label{eq:Haar_2nd}
    \int_{U\in\mathcal{U}}d\mu_{\mathrm{Haar}}(U) U^\dagger|b\rangle \langle b|U  \langle b|U O_1 U^\dagger|b\rangle 
    &= \frac{O_1+\mathds{1}\Tr{O_1}}{2^n(2^n+1)}. 
    \\
    \label{eq:Haar_3rd}
    \int_{U\in\mathcal{U}}d\mu_{\mathrm{Haar}}(U) U^\dag\ket{b}\bra{b}U \bra{b}U O_2 
    U^\dag\ket{b}\bra{b}U O_3 U^\dag\ket{b}
    &=\frac{\mathds{1}\,\Tr[O_2O_3]+O_2O_3+O_3O_2}{2^n(2^{n}+1)(2^{n}+2)}.
\end{align}
Here, $O_1$ is an arbitrary Hermitian operator and $O_2$ and $O_3$ are traceless operators.

Using the second-moment Haar integral given in Eq.~\eqref{eq:Haar_2nd}, we find that
\begin{align}
    \sum_{b\in \{0, 1\}^{n}}\int_{U\in\mathcal{U}}d\mu_{\mathrm{Haar}}(U) \langle b|U\rho U^\dagger|b\rangle U^\dagger|b\rangle \langle b|U  
    = \frac{\rho+\mathds{1}\Tr{\rho}}{(2^n+1)}. 
\end{align} 
This leads to the following inverse quantum channel for $U\in\mathcal{U}_{k\geq2}$:
\begin{equation}
    \mathcal{M}(\rho) = \frac{\rho+\mathds{1}\Tr{\rho}}{(2^n+1)} \quad \Longrightarrow \quad  \mathcal{M}^{-1}(X) = (2^n +1)X - \mathds{1}\Tr{X}, 
\end{equation} 
where we use the fact that the channel is unital, $\mathcal{M}(\mathds{1})=\mathds{1}$, and $X$ is a Hermitian operator of dimension $2^n\times 2^n$. As such, the classical shadows~\eqref{eq:shadow} for random $n$-qubit Clifford measurements of the state $\rho$ are given by 
\begin{equation}
\label{eq:Clshaddows}
    \hat{\rho}_{U,b} = (2^n +1)U^\dagger|b\rangle \langle b|U - \mathds{1}.
\end{equation}

\subsection{Computing expectation values with classical shadows}

Having defined the classical shadows and shown how to construct them from outcomes of randomized measurements, we can now look into the problem of computing expectation values. Suppose that we are interested in expectation values of the form $\Tr{\rho O_j}$, for the set of Hermitian operators $\{O_j\}_{j=1}^{M}$. In this subsection, we show that we can compute these expectation values by taking the classical shadows, $\hat{\rho}_{U,b}$, as estimators for the density matrix. To keep this review concise, we only focus on the randomized Clifford measurements, for which the shadows are given in Eq.~\eqref{eq:Clshaddows}.

In order to show that we can use $\hat{\rho}_{U,b}$ as an estimator for the expectation values of $\rho$, we need to compute the average, $\mathbb{E}\left[\Tr{\hat{\rho}_{U,b}O_j}\right]$, and the variance, $\text{Var}\left[\Tr{\hat{\rho}_{U,b}O_j}\right]$, with respect to the randomized measurement process. As the estimators $\hat{\rho}_{U,b}$ are obtained from randomized Clifford measurements, we can compute these averages by again exploiting the fact that $U\in\mathcal{C}l(2^n)$ is a unitary 3-design. 
First, by using the linearity of the trace and Eq.~\eqref{eq:recrho}, we find that
\begin{align}
    \mathbb{E}\left[\Tr{\hat{\rho}_{U,b}O_j}\right] = \Tr \mathbb{E}[\hat{\rho}_{U,b}]O_j = \Tr{\rho O_j}
\end{align} 
for the average of the shadow expectation value of observable $O_j$.
For the computation of the variance, we note that it only depends on the traceless part, $O_j^0=O_j-\frac{\mathds{1}\Tr{O_j}}{2^n}$, of $O_j$. This can be seen from
\begin{align}
    \text{Var}\left[\Tr{\hat{\rho}_{U,b} O_j} \right] &= \mathbb{E}\left[ (\Tr{\hat{\rho}_{U,b}O_j} -\mathbb{E}\left[\Tr{\hat{\rho}_{U,b}O_j} \right])^2\right]
    = \mathbb{E}[(\Tr{\hat{\rho}_{U,b}O_j^0})^2] - \left(\Tr{\rho_{U,b} O^0_j} \right)^2.
\end{align} 
This important observation simplifies the variance calculation enormously, as we now only need to compute one Haar integral.
Focusing on the first term in the variance, we obtain 
\begin{align}
    \mathbb{E}\big[\big(\Tr{\hat{\rho}_{U,b}O_j^0}\big)^2\big] 
    &= \mathbb{E}\big[\big(\Tr\mathcal{M}^{-1}(U^\dagger \ket{b}\bra{b}U) O_j^0\big)^2 \big] \\
    &= (2^n+1)^2\sum_{b\in\{0, 1\}^n}\int_{U\in\mathcal{U}}d\mu_{\mathrm{Haar}}(U)\Tr{\rho U^\dagger|b\rangle \langle b|U}\bra{b}UO_j^0U^\dagger\ket{b}^2 \label{eq:line2} \\ 
    &= \frac{2^n+1}{2^n+2}\left(\Tr{(O_j^0)^2} +2\Tr{\rho (O_j^0)^2}\right).\label{eq:line3}
\end{align} 
Here, we use the identity for the third-moment Haar integral~\eqref{eq:Haar_3rd} for traceless operators after swapping the trace and integral in Eq.~\eqref{eq:line2}. Subsequently, we perform the sum over all bit strings to obtain the result in Eq.~\eqref{eq:line3}. The variance is thus equal to 
\begin{equation}\label{eq:cliffshadowvar}
     \text{Var}\left[\Tr{\hat{\rho}_{U,b} O_j} \right] =  \frac{2^n+1}{2^n+2}\left(\Tr{(O_j^0)^2} +2\Tr{\rho (O_j^0)^2}\right) - \left(\Tr{\rho O_j^0} \right)^2. 
\end{equation}

\subsection{Predicting \texorpdfstring{$M$}{M} observables with \texorpdfstring{$\mathcal{O}(\log{M})$}{O(log(M))} classical shadows}

For the final part of our review of the classical-shadow-tomography algorithm, we show that we only need $\mathcal{O}(\log{M})$ independent classical shadows, $\hat{\rho}_{U,b}$, to predict $M$ expectation values $\{\Tr{\rho O_j}\}_{j=1}^{M}$. To simplify the notation, we replace the $U$ and $b$ subscripts with an index; i.e., now $\hat{\rho}_i$ denotes the $i$th shadow. We follow the approach laid out in Ref.~\cite{Huang2020} closely and perform a median-of-means estimation. The median-of-means estimator of $\Tr{\rho O_j}$ is defined as 
\begin{equation}
    \mu_j(N, K) = \text{median}\left[\mu_j^1(N), \mu_j^2(N), \dots, \mu_j^K(N) \right],
\end{equation} 
with
\begin{equation}
    \mu_j^k(N) = \frac{1}{N}\sum_{i=N(k-1)+1}^{Nk} \Tr{\hat{\rho}_i O_j}.
\end{equation} 
This means that we construct $n_{\mathrm{shadows}}=NK$ independent shadows $\hat{\rho}_i$ in total, divide them into $K$ subsets of size $N$, compute the mean of each subset, and finally take the median of the subset means. 

Although this procedure is slightly more complicated than computing the mean directly on all the shadows, there are scenarios in which the median-of-means estimator is a more robust estimator than the standard mean estimator (see, e.g., Ref.~\cite{Lerasle2019}). Moreover, results are known for how many samples (independent estimators) are required to predict a function up to a certain accuracy. Specifically, for a random variable $X$ with finite sample variance $\sigma^2$, the median-of-means estimator, $\tilde{\mu}(N,K)$, has the following failure probability~\cite{Lerasle2019, Chen2020}:
\begin{equation}
\label{eq:MoM_ineq}
    \mathrm{Pr}\left[|\tilde{\mu}(N,K) - \mathbb{E}\left[X\right]|\geq \epsilon\right]\leq e^{-2K\left(\frac{1}{2} - \frac{\sigma^2}{N\epsilon^2} \right)^2},
\end{equation} 
where the number of shadows in each subset needs to satisfy $N>2\sigma^2/\epsilon^2$ for this inequality to hold. Then, by setting $K=(9/2)\log(M/\delta)$ and $N=6\sigma^2/\epsilon^2$, we find that
\begin{equation}
\label{eq:MoM_ineq_optimal}
    \mathrm{Pr}\left[|\tilde{\mu}(N,K) - \mathbb{E}\left[X\right]|\geq \epsilon\right]
    \leq  \frac{\delta}{M}.
\end{equation}
The values of the constants $N$ and $K$ are found by minimizing the number of shadows for a fixed upper bound $\delta/M$ and with $N>2\sigma^2/\epsilon^2$. In this way, we obtain a number of shadows $n_{\mathrm{shadows}}=NK$ smaller than the (reserved) one found in Ref.~\cite{Huang2020}.

Now, by applying a median-of-means estimation to the problem of computing $M$ observables $O_j$, we find that
\begin{equation}
    \text{Pr}\left[|\mu_j(N,K) - \Tr{\rho O_j}|\geq \epsilon\right]\leq \frac{\delta}{M}.
\end{equation} 
As this is the failure probability for each individual observable $O_j$, the failure probability of all of them together (i.e., that at least one observable fails) follows from Boole's inequality,
\begin{equation}
    \text{Pr}\Big[ \max_{j\in\{1,\dots,M\}}|\mu_j(N,K) - \Tr{\rho O_j}|\geq \epsilon\Big]\leq \delta.
\end{equation} 
From this, we arrive at the following theorem~\cite{Huang2020}. 

\begin{theorem}[Number of shadows for $M$ observables] 
\label{thm:2}
Given a random measurement process with unitary ensemble $\mathcal{U}$, $M$ Hermitian operators $O_j$, and accuracy parameters $\epsilon, \delta\in[0,1]$, in total 
\begin{equation}\label{eq:numshadows}
    n_{\mathrm{shadows}}=NK = \frac{27\log(M/\delta)}{\epsilon^2}\max_j\left[\sigma_j^2\right]
\end{equation} 
classical shadows $\hat{\rho}$, with $\sigma_j^2$ the shadow variance (see Eq.~\eqref{eq:cliffshadowvar} for Cliffords) for observable $O_j$, are sufficient to estimate all $M$ expectation values, $\Tr \rho O_j$, with a success probability of
\begin{equation}
    \mathrm{Pr}\Big[\max_{j\in\{1,\dots,M\}}|\mu_j(N,K) - \Tr{\rho O_j}|\leq \epsilon\Big]\geq 1-\delta.
\end{equation} 
\end{theorem}

This means that in total we need  $\mathcal{O}\left(\log{M}\right)$ randomized measurements. Importantly, the mean squared error for observable $O_j$, $\sigma_j^2$ depends on the operator that one tries to measure, and which specific unitary ensemble is used. As a final remark, note that the constants in Eq.~\eqref{eq:numshadows} result from bounding the failure probability. This means that it is a worst-case scenario (upper bound) and in practice one could likely get away with fewer shadows.

\section{Proof of exponentially vanishing failure probability of thermal pure quantum states}\label{app:TPQ}

In this appendix we show that the random state 
\begin{align}\label{eq:tpq}
    \ket{\psi_\beta} = \frac{e^{-\beta H/2}U\ket{0}}{\sqrt{\bra{0}U^\dagger e^{-\beta H} U\ket{0}}},
\end{align}
where the unitary $U\in \mathcal{U}_{k\geq2}$ is at least a quantum 2-design, satisfies Eq.~\eqref{eq:purethermalstate} and therefore is a thermal pure quantum (TPQ) state. We start by deriving the expectation value and variance of an observable $O$ in the state $\ket{\psi_\beta}$, for which the results are given in Eqs.~\eqref{eq:etherm} and \eqref{eq:var}. Afterward, we apply Markov's inequality to prove the exponentially vanishing failure probability. 

\subsection{Deriving the variance and expectation value of the thermal pure quantum state}

First, we note that the variance of the normalization constant, $\bra{0}U^\dagger e^{-\beta H} U\ket{0}$, of the TPQ state, $\ket{\psi_\beta}$, is exponentially suppressed. This can be derived by evaluating the variance explicitly and invoking the Haar-integral identities in Eqs.~\eqref{eq:Haar_1st} and \eqref{eq:Haar_2nd}:
\begin{align}
    \mathrm{Var}\left[\bra{0}U^\dagger e^{-\beta H} U\ket{0}\right] 
    =\mathbb{E}_U\big[\bra{0}U^\dagger e^{-\beta H} U\ket{0}^2\big]
    -\big(\mathbb{E}_U[\bra{0}U^\dagger e^{-\beta H} U\ket{0}]\big)^2
    = \frac{2^n\Tr{e^{-2\beta H}} - (\Tr{e^{-\beta H}})^2}{2^{2n}(2^n+1)}.
\end{align} 
Since $0<\Tr{e^{-2\beta H}}\leq 2^n$ for a positive semidefinite $H$\footnote{Note that the spectrum of $H$ can be shifted to make it positive semidefinite. This does not affect our results, as the shift in energy cancels between the numerator and denominator in Eq.~\eqref{eq:tpq}.}, and for all finite $\beta>0$, this variance will always be exponentially small in the system size, $n$. 

In the following, we use the shorthand notation $f\equiv \bra{0}U^\dagger e^{-\beta H/2} O e^{-\beta H/2}U\ket{0}$ and $g \equiv\bra{0}U^\dagger e^{-\beta H} U\ket{0}$. 
As $\mathrm{Var}\left[g\right]$ is small, we expand the expectation value of $O$ with a multivariate Taylor expansion up to first order in $\mathrm{Var}\left[g\right]$:
\begin{align}
\label{eq:etherm2}
    \mathbb{E}\left[\bra{\psi_\beta}O\ket{\psi_\beta}\right] = \mathbb{E}\left[\frac{f}{g}\right] \approx 
    \frac{\mathbb{E}\left[f\right]}{\mathbb{E}\left[g\right]} - \frac{\mathrm{Cov}\left[f,g\right]}{\mathbb{E}\left[g\right]^2} + \frac{\mathbb{E}\left[f\right]}{\mathbb{E}\left[g\right]^3}\mathrm{Var}\left[g\right],
\end{align} 
where $\mathrm{Cov}[f,g]$ denotes the covariance between $f$ and $g$, defined by $\mathrm{Cov}[f,g]\equiv \mathbb{E}[fg] -\mathbb{E}[f]\mathbb{E}[g]$. Note that here we do not explicitly write the higher-order terms in the Taylor expansion. These higher-order terms need to be evaluated, or bounded, to turn the approximate equality into an equality.
Now, by evaluating each of the terms in this Taylor expansion with the Haar-integral identities in Eqs.~\eqref{eq:Haar_1st} and \eqref{eq:Haar_2nd}, we arrive at Eq.~\eqref{eq:etherm}:
\begin{equation}
   \mathbb{E}\left[\bra{\psi_\beta}O\ket{\psi_\beta}\right] \approx  \Tr{\rho_\beta O} + \Tr{\rho_\beta^2}\left(\Tr{\rho_\beta O} - \Tr{\rho_{2\beta} O} \right).
\end{equation}

For the variance of the expectation value, we exploit a similar multivariate Taylor expansion, 
\begin{equation}
    \mathrm{Var}\left[\frac{f}{g}\right]\approx \frac{\mathrm{Var}\left[f\right]}{\mathbb{E}\left[g\right]^2} - \frac{2\mathbb{E}\left[f\right]}{\mathbb{E}\left[g\right]^3}\mathrm{Cov}\left[f, g\right] + \frac{\mathbb{E}\left[f\right]^2}{\mathbb{E}\left[g\right]^4}\mathrm{Var}\left[g\right],
\end{equation}
and obtain Eq.~\eqref{eq:var} after the evaluation of the Haar integrals 
\begin{equation}\label{eq:varterm}
   \mathrm{Var}\left[ \bra{\psi_\beta}O\ket{\psi_\beta}\right] \approx \Tr{\rho_\beta^2}\left(\frac{\Tr(Oe^{-\beta H})^2}{\Tr{e^{-2 \beta H}}} -2 \Tr{\rho_\beta O}\Tr{\rho_{2\beta} O} + (\Tr{\rho_\beta O})^2\right).
\end{equation} 

\subsection{Application of Markov's inequality}

In order to show that $\ket{\psi_\beta}$ satisfies Eq.~(\ref{eq:purethermalstate}), we start from the following Markov inequality:
\begin{align}
   &\text{Pr}\left[|\bra{\psi_\beta}O\ket{\psi_\beta} - \Tr{\rho_\beta O}|\geq \epsilon\right]\leq \frac{\mathbb{E}_U\left[\bra{\psi_\beta}O\ket{\psi_\beta} - \Tr{\rho_\beta O}\right]^2}{\epsilon^2}
\end{align} 
Here, $\mathbb{E}_U$ denotes the ensemble average with respect to the $n$-qubit Clifford group $U\in\mathcal{C}l(2^n)$. This average can be computed by using the unitary 2-design property of $U$ and writing it in terms of Haar integrals. After evaluating the integrals, this results in
\begin{align} 
    \mathbb{E}_U\left[\bra{\psi_\beta}O\ket{\psi_\beta} - \Tr{\rho_\beta O}\right]^2 
    \approx \Tr{\rho_\beta^2}\left(\frac{\Tr{(Oe^{-\beta H})^2}}{\Tr{e^{-2 \beta H}}} -2 \Tr{\rho_\beta O}\Tr{\rho_{2\beta} O} + (\Tr{\rho_\beta O})^2\right).
\end{align} 
The terms that multiply the purity, $\Tr{\rho_\beta^2}$, in this expression, have the form of expectation values, which can be upper bounded by the spectral norm, $\lVert O\rVert^2$. For the first term, $\frac{\Tr{(Oe^{-\beta H})^2}}{\Tr{e^{-2 \beta H}}}$, this can be proven as follows. Inserting the complete set of eigenstates $\{\ket{\lambda}\}$ of $H$, i.e., $H\ket{\lambda}=\lambda\ket{\lambda}$, we have
\begin{align}
\begin{split}
    \frac{\Tr{(Oe^{-\beta H})^2}}{\Tr{e^{-2 \beta H}}}
    &= \frac{\sum_{\lambda,\tilde{\lambda}}\Tr[Oe^{-\beta H}\ket{\lambda}\bra{\lambda} Oe^{-\beta H}\ket{\tilde{\lambda}}\bra{\tilde{\lambda}}]}{\Tr{e^{-2 \beta H}}} 
    =\frac{\sum_{\lambda,\tilde{\lambda}}
    e^{-\beta \lambda}e^{-\beta \tilde{\lambda}}
    |\bra{\lambda} O\ket{\tilde{\lambda}}|^2}{\Tr{e^{-2 \beta H}}}
    \\
    &\le\frac{1}{\Tr{e^{-2 \beta H}}}\sum_{\lambda,\tilde{\lambda}}
    \frac{e^{-2\beta \lambda}+e^{-2\beta \tilde{\lambda}}}{2}
    |\bra{\lambda} O\ket{\tilde{\lambda}}|^2
    =\Tr[\rho_{2\beta} O^2]
    \le \lVert O^2\rVert 
    \le \lVert O\rVert^2,
\end{split}
\end{align}
where we use the inequality of arithmetic and geometric means, $\sqrt{xy}\le(x+y)/2$ for $x,y\in\mathbb{R}_{\ge0}$.

We thus arrive at 
\begin{equation}
   \text{Pr}\left[|\bra{\psi_\beta}O\ket{\psi_\beta} - \Tr{\rho_\beta O}|\geq \epsilon\right] \lesssim 4 \frac{\lVert O\rVert^2 \Tr{\rho_\beta^2}}{\epsilon^2}, 
\end{equation} 
which means that $\ket{\psi_\beta}$ satisfies Eq.~\eqref{eq:purethermalstate} with $C_\epsilon \approx 4\lVert O\rVert^2/\epsilon^2$ and $e^{-\alpha n}=\Tr{\rho_\beta^2}$. The fact that $\Tr{\rho_{\beta}^2}$ can be written as an exponential is explained below Eq.~\eqref{eq:purity} in the main text.

\section{Proof of the main result -- The required number of shadows for predicting \texorpdfstring{$M$}{M} thermal expectation values}\label{app:proof}

In this appendix, we provide the detailed proof of Theorem~\ref{thm:1}. We construct classical shadows of TPQ states (pure thermal shadows) by using two independent random unitaries. The first unitary, $U\sim \mathcal{C}l(2^n)$, is followed by an application of $e^{-\beta H/2}$ to create a TPQ state, $\ket{\psi_\beta}$. The second unitary, $V$, combined with the measurement outcomes in the form of a bit string, $\ket{b}$, is used to construct the corresponding classical shadow, $\hat{\eta}_{V,b}=\mathcal{M}^{-1}(V^\dag\ket{b}\bra{b}V)$. We consider the cases where $V$ is drawn from $\mathcal{C}l(2^n)$, a random $n$-qubit Clifford measurement, and $\mathcal{C}l(2)^{\otimes n}$, a random Pauli measurement, respectively. We start by deriving the mean squared errors of the thermal shadows, Eqs.~\eqref{eq:shadowvar} and \eqref{eq:shadowvar_pauli}, for both measurement protocols. Then, from a direct application of Theorem~\ref{thm:2}, we obtain Theorem~\ref{thm:1}, which completes the proof.

\subsection{Random Clifford measurements}

The random-Clifford-measurement protocol consists of the application of  a Clifford circuit $V\sim\mathcal{C}l(2^n)$ to the TPQ state, $\ket{\psi_\beta}$, followed by a computational-basis measurement with bit-string outcome $\ket{b}$. We first compute the average, $\mathbb{E}\left[\Tr{\hat{\eta}_{V,b}O}\right]$, and then the mean squared error, $\sigma^2$, with respect to the random Clifford measurement process.
For the average, we find that
\begin{align}
\label{eq:shadowavg}
    \mathbb{E}\left[\Tr{\hat{\eta}_{V,b}O}\right] =\mathbb{E}_{U}\mathbb{E}_{V,b}\left[\Tr{\hat{\eta}_{V,b}O}\right] = \mathbb{E}_{U}\left[\Tr{\mathbb{E}_{V,b}\left[\hat{\eta}_{V,b}\right]O}\right]
    = \mathbb{E}_{U}[\bra{\psi_\beta} O\ket{\psi_\beta}]
    \approx \Tr{\rho_\beta O} +\mathcal{O}(e^{-n}),
\end{align} 
where we use the fact that the averaged shadow reproduces the TPQ state, $\mathbb{E}_{V,b}\left[\hat{\eta}_{V,b}\right]=\ket{\psi_\beta}\bra{\psi_\beta}$. 

In order to compute the mean squared error, we first write it in terms of the variance of the thermal-shadow expectation value: 
\begin{equation}
\label{eq:thermal_shadow_sigma}
    \sigma^2 = \mathbb{E}_U\mathbb{E}_{V,b}[(\Tr\hat{\eta}_{V,b}{O}-\Tr{\rho_\beta O})^2]
    = \mathrm{Var}[\Tr{\hat{\eta}_{V,b}O}]+ (\mathbb{E}_U\mathbb{E}_{V,b}[\Tr\hat{\eta}_{V,b}{O}]-\Tr{\rho_\beta O})^2
    \approx \mathrm{Var}[\Tr{\hat{\eta}_{V,b}O}] + \mathcal{O}(e^{-n}).
\end{equation} 
The thermal-shadow variance can be computed from 
\begin{align}
\label{eq:TPQshadow_var}
	\mathrm{Var}[\Tr{\hat{\eta}_{V,b}O}] &= \mathbb{E}_U\mathbb{E}_{V,b}\big[\big(\Tr{\hat{\eta}_{V,b}O} - \mathbb{E}_U\mathbb{E}_{V,b}[\Tr{\hat{\eta}_{V,b} O}]\big)^2\big]
	= \mathbb{E}_U\mathbb{E}_{V,b}\big[(\Tr{\hat{\eta}_{V,b}O_0})^2\big] 
	- \big(\mathbb{E}_U\mathbb{E}_{V,b}[\Tr{\hat{\eta}_{V,b} O_0}]\big)^2,
\end{align}
where $O_0$ is the traceless part of the observable $O$.
Focusing on the first term and invoking Eq.~\eqref{eq:line3} for the random measurement process, we obtain
\begin{align}
    \mathbb{E}_{U}\mathbb{E}_{V,b}\left[ (\Tr{\hat{\eta}_{V,b}O_0})^2\right] 
    = \frac{2^n+1}{2^n+2}\mathbb{E}_{U}\big[\Tr{O_0^2} +2\bra{\psi_\beta} O_0^2\ket{\psi_\beta}\big] 
    \approx \frac{2^n+1}{2^n+2}\left(\Tr{O_0^2} +2\Tr{\rho_\beta O_0^2}\right) + \mathcal{O}(e^{-n}).
\end{align} 
From this, we find that the variance is equal to
\begin{equation}
\label{eq:shadow_var_Clifford}
     \text{Var}\left[\Tr{\hat{\eta}_{V,b} O} \right] 
     \approx
     \frac{2^n+1}{2^n+2}\left(\Tr{O_0^2} +2\Tr{\rho_\beta O_0^2}\right) - (\Tr{\rho_\beta O_0})^2 +\mathcal{O}(e^{-n}). 
\end{equation} 
The combination of this expression with Eq.~\eqref{eq:thermal_shadow_sigma} then results in Eq.~\eqref{eq:shadowvar} for the mean squared error. 

We note that the mean squared error of the thermal shadows is, up to an exponential correction, equal to the $\sigma^2$ that one obtains by plugging $\rho=\rho_\beta$ in Eq.~\eqref{eq:cliffshadowvar}. As such, for sufficiently large systems, shadows obtained from randomly measuring the TPQ state perform approximately as well as shadows obtained from measuring the true Gibbs state.

\subsection{Random Pauli measurements}

Random Pauli measurements can be performed by applying a tensor product of random single-qubit Clifford gates, $V=\bigotimes_{i=1}^{n}V_i$, to the TPQ state, $\ket{\psi_\beta}$, followed again by measurements in the computational basis with bit-string outcomes, $\ket{b}=\bigotimes_{i=1}^n\ket{b_i}$. 
For these random Pauli measurements, the quantum channel $\mathcal{M}$ [Eq.~\eqref{eq:mchannel}] is given by the depolarizing channel, $(D_{1/3})^{\otimes n}$ (for the derivation, see, e.g., Ref.~\cite{Huang2020}). This means that the inverse channel is given by $\mathcal{M}^{-1}(X)=(D_{1/3}^{-1})^{\otimes n}(X)$ and the classical shadow can be constructed from,
\begin{align}\label{eq:pshadow}
    \hat{\eta}_{V,b}
    =\bigotimes_{i=1}^{n} D_{1/3}^{-1}(V_i^\dag\ket{b_i}\bra{b_i}V_i)
    =\bigotimes_{i=1}^{n} (3\,V_i^\dag\ket{b_i}\bra{b_i}V_i - \mathds{1}_i).
\end{align}
These shadows are just product states and hence can be stored and efficiently computed classically. From the definition of classical shadows [Eq.~\eqref{eq:recrho}], the shadows satisfy $\mathbb{E}\left[\hat{\eta}_{V,b}\right]=\ket{\psi_\beta}\bra{\psi_\beta}$. For this reason, the application of Eq.~\eqref{eq:shadowavg}, results in the average 
\begin{equation}
    \mathbb{E}\left[\Tr\hat{\eta}_{V,b}O\right] = \mathbb{E}_U\mathbb{E}_{V,b}[\Tr\hat{\eta}_{V,b}O] \approx \Tr{\rho_\beta O} + \mathcal{O}(e^{-n}),
\end{equation} 
for the expectation value of the Pauli shadows in Eq.~\eqref{eq:pshadow}. Thus, shadows from randomized Pauli measurements can, like Clifford measurements, be used as estimators for Gibbs-state expectation values.

For the mean squared error, we use Eq.~\eqref{eq:thermal_shadow_sigma} again and evaluate the first term in the variance [Eq.~\eqref{eq:TPQshadow_var}] for the shadows constructed from Pauli measurements.
For this we assume that we are interested in an observable $O$, which is a $k$-local Pauli operator, $O=P_1\otimes\dots\otimes P_k\otimes I^{\otimes (n-k)}$. For more general observables, our calculation below can be modified by using the results presented in the appendix of Ref.~\cite{Huang2020}. We find that the first term in the variance is given by
\begin{align}
\label{eq:shadow_var_Pauli}
	\mathbb{E}_U\mathbb{E}_{V,b}\big[(\Tr{\hat{\eta}_{V,b}O_0})^2\big]
	&= \mathbb{E}_{U}\mathbb{E}_{V}\sum_{b\in\{0,1\}^n} |\bra{\psi_\beta}V\ket{b}|^2
	\Tr[O_0\,(\mathcal{D}^{-1}_{1/3})^{\otimes n}(V^\dag\ket{b}\bra{b}V)]^2
	\\
	&= \Tr\Big[\ket{\psi_\beta}\bra{\psi_\beta} \bigotimes_{j=1}^{k} 
	\mathbb{E}_{V_j}\sum_{b_j\in\{0,1\}} V_j^\dag\ket{b_j}\bra{b_j}V_j 
	(3\bra{b_j}V_j P_jV_j^\dag\ket{b_j})^2\Big]
	= 3^k.
\end{align} 
Here, we use the second-moment Haar integral [Eq.~\eqref{eq:Haar_2nd}] for a single qubit to compute the average over $\mathbb{E}_{V_j}$ and the fact that the contribution from $I^{\otimes (n-k)}$ evaluates to 1. The combination of this with the second term in Eq.~\eqref{eq:TPQshadow_var}, and with Eq.~\eqref{eq:thermal_shadow_sigma}, leads to Eq.~\eqref{eq:shadowvar_pauli}.

\subsection{Proof of Theorem~\ref{thm:1}}

A direct application of Theorem~\ref{thm:2} proves Theorem~\ref{thm:1}. The median-of-means estimator of the Gibbs-state expectation values, $\Tr{\rho_\beta O_j}$, by the thermal shadows $\{\hat{\eta}_{V,b}\}$ is now defined as 
\begin{align}
    \mu_j(N, K) = \text{median}\left[\mu_j^1(N), \mu_j^2(N), \dots, \mu_j^K(N) \right],
    \qquad
    \mu_j^i(N) = \frac{1}{N}\sum_{\ell=N(i-1)+1}^{Ni} \Tr\hat{\eta}_{V,b}{O_j}.
\end{align} 
Then, Theorem~\ref{thm:2} holds with $\sigma_j^2$ replaced by Eq.~\eqref{eq:thermal_shadow_sigma} for the mean squared error of observable $O_j$. These errors are for random Clifford and Pauli measurements given by Eqs.~\eqref{eq:shadowvar} and \eqref{eq:shadowvar_pauli}. This completes the proof of Theorem~\ref{thm:1}.

\section{Quantum signal processing and the quantum eigenvalue transformation}
\label{app:qsp}

In this appendix, we review quantum signal processing (QSP), a framework for performing matrix arithmetic on quantum computers. QSP produces a degree-$d$ polynomial $f(x)$ for a real scalar value $x$ such that~\cite{Low_2016,Low_2017,Low_2019}:
\begin{enumerate}[i]
    \item $f$ has parity-$(d \mathrm{\ mod\ }2)$ 
    \item $|f(x)|\le1$ for all $x\in[-1,1]$
    \item  $|f(x)|\ge1$ for all $x\in(-\infty,1]\cup(1,\infty]$ and 
    \item $f(i x)f^\ast(i x)\ge1$ for all $x\in\mathbb{R}$ if $d$ is even
\end{enumerate} 
Indeed, there exists a set of real parameters $\{\phi_1,\dots,\phi_d\}$ that produces the following two-dimensional unitary matrix (Corollary 8 in Ref.~\cite{Gilyen_2019}):
\begin{align}
\label{eq:qsp_unitary}
	U_{\boldsymbol{\phi}} := \prod_{k=1}^{d}S(\phi_k)W(x) = 
	\begin{pmatrix}
		f(x) & \ast \\
		\ast & \ast
	\end{pmatrix},
\end{align}
where a signal-processing rotation operator $S(\phi)$ and a signal rotation operator $W(x)$ are, respectively, given by
\begin{align}
	S(\phi) = e^{i\phi \sigma_z},
	\qquad
	W(x) = X e^{-i \arcsin(x)\sigma_y} =
	\begin{pmatrix}
		x & \sqrt{1-x^2} \\
		\sqrt{1-x^2} & -x
	\end{pmatrix}.
 \label{eq:sig_rotation_op}
\end{align} 
From this, a real polynomial $f_\Re(x)$ of degree-$d$ is obtained by
\begin{align}
\label{eq:real_qsp}
	f_\Re(x) 
	= (\bra{+}\otimes \bra{0})(\mathds{1}\otimes U_{\boldsymbol{\phi}} + \ket{1}\bra{1}\otimes U_{-2\boldsymbol{\phi}})(\ket{+}\otimes \ket{0}) 
	= \frac{f(x)+f^\ast(x)}{2}.
\end{align}

QSP can be promoted to a quantum eigenvalue transformation (QET), which applies a polynomial transformation to a Hermitian matrix in input~\cite{Martyn_2021_grand}.
In this approach, the QSP unitary~\eqref{eq:qsp_unitary} is realized in a qubitized subspace, i.e., a two-dimensional subspace of a larger Hilbert space that is identified with a single-qubit space.
We begin by block encoding the input matrix $H$ on a primary $n$-qubit system, i.e., embedding $H$ into a unitary matrix $\mathcal{W}$ by adding a $m$-qubit ancillary register. For any state $\ket{\psi}$ of the primary system, $\mathcal{W}$ is defined such that
\begin{align}
\label{eq:block_encoding}
	\mathcal{W} \ket{0^m}\otimes\ket{\psi}=\ket{0^m}\otimes \frac{H}{a}\ket{\psi} + \ket{\Psi^\perp},
\end{align}
where $\ket{\Psi^\perp}$ satisfies $(\bra{0^m}\otimes \mathds{1}) \ket{\Psi^\perp}=0$ and $a$ is some constant obeying $a\ge\lVert H\rVert$.

To construct such a block encoding, we employ the linear combination of unitaries (LCU) method~\cite{Childs2012}. The LCU provides a way to block encode $H$ when it is expressed as a linear combination of unitaries $\{P_k\}_{k=1}^{K}$, $H=\sum_{k=1}^K a_k P_k$ with $a_k\in\mathbb{R}$. Following the main text, we take the unitary operators to be Pauli operators $P_k\in\{\mathds{1}, \sigma_x, \sigma_y, \sigma_z\}^{\otimes n}$.
The LCU consists of two unitary operators: 
\begin{enumerate}[i]
    \item a unitary operator $A$ acting on the ancillary register with $m=\log K$ such that
 $A\ket{0^m} = \frac{1}{\sqrt{a}}\sum_k\sqrt{a_k}\ket{k}$
with $a=\sum_k a_k$ and
\item a controlled unitary operator $B = \sum_{k=1}^{K}\mathrm{sign}(a_k)\ket{k}\bra{k}\otimes P_k$ with the sign function, $\mathrm{sign}(a)=+1(-1)$ for $a\ge0(a<0)$~\footnote{Alternatively, the sign function $\mathrm{sign}(a_k)$ could be absorbed into the unitary operator $P_k$.}
\end{enumerate}
Then, one can readily show that $\mathcal{W}=A^\dag BA$ gives a block encoding of $H$, i.e., it satisfies Eq.~\eqref{eq:block_encoding}.

In order to clarify the connection with QSP, we expand the input state, $\ket{\psi}=\sum_\lambda c_\lambda\ket{\lambda}$, in the eigenbasis of $H$, $(H/a)\ket{\lambda}=\lambda\ket{\lambda}$.
In each eigensubspace spanned by the basis vectors, 
\begin{align}
\label{eq:qubitized_basis}
	\left\{\ket{0^m}\otimes\ket{\lambda}, \frac{(\mathcal{W}-\lambda)\ket{0^m}}{\sqrt{1-\lambda^2}}\otimes\ket{\lambda}\right\},
\end{align}
the block encoding $\mathcal{W}$ has the matrix representation
\begin{align}
	\begin{pmatrix}
		\lambda & \sqrt{1-\lambda^2} \\
		\sqrt{1-\lambda^2} & -\lambda
	\end{pmatrix}.
\end{align}
Therefore, the block encoding $\mathcal{W}$ can be identified with the signal-processing operator $W$ [Eq.~\eqref{eq:sig_rotation_op}] in the qubitized space spanned by Eq.~\eqref{eq:qubitized_basis}.
The signal-processing rotation operator $\mathcal{S}(\phi)$ in the qubitized space is constructed as
\begin{align}
	\mathcal{S}(\phi) = (\bra{0}\otimes \mathds{1}^{\otimes m})\,C_{\ket{0}}\text{NOT}\,(e^{-i\phi \sigma_z}\otimes \mathds{1}^{\otimes m})\,C_{\ket{0}}\text{NOT}\,(\ket{0}\otimes \mathds{1}^{\otimes m}),
\end{align}
where we have added a single-qubit ancillary register and defined $C_{\ket{0}}\text{NOT}=\sigma_x\otimes\ket{0^m}\bra{0^m}+\mathds{1}\otimes(\mathds{1}^{\otimes m}-\ket{0^m}\bra{0^m})$. This operator $\mathcal{S}(\phi)$ acts trivially on the primary system.
Combining them, we find that the following sequential application of unitaries,
\begin{align}
\label{eq:qet_unitary}
	\mathcal{U}_{\boldsymbol{\phi}} := \prod_{k=1}^{d} \big(\mathcal{S}(\phi_d)\mathcal{W}(H)\big) = 
	\bigoplus_\lambda
	\begin{pmatrix}
		f(\lambda) & \ast \\
		\ast & \ast
	\end{pmatrix}
	= 
	\begin{pmatrix}
		f(H) & \ast \\
		\ast & \ast
	\end{pmatrix},
\end{align}
yields the polynomial $f(H)$ in the top-left block ($\ket{0^m}\bra{0^m}$ component) of the unitary~$\mathcal{U}_{\boldsymbol{\phi}}$.
The corresponding real polynomial $f_{\Re}(H)$ is obtained by using Eq.~\eqref{eq:real_qsp}.

\section{Additional numerical experiments}\label{app:add_num}

In this appendix, we present additional numerical experiments that support our claims in the main text. We first show results for the efficiency of the TPQ states. We then explain how to train a quantum Boltzmann machine (QBM) with pure thermal shadows.

\subsection{Variance of thermal pure quantum states}

\begin{figure}[ht]\centering
    \includegraphics[width=0.92\columnwidth]{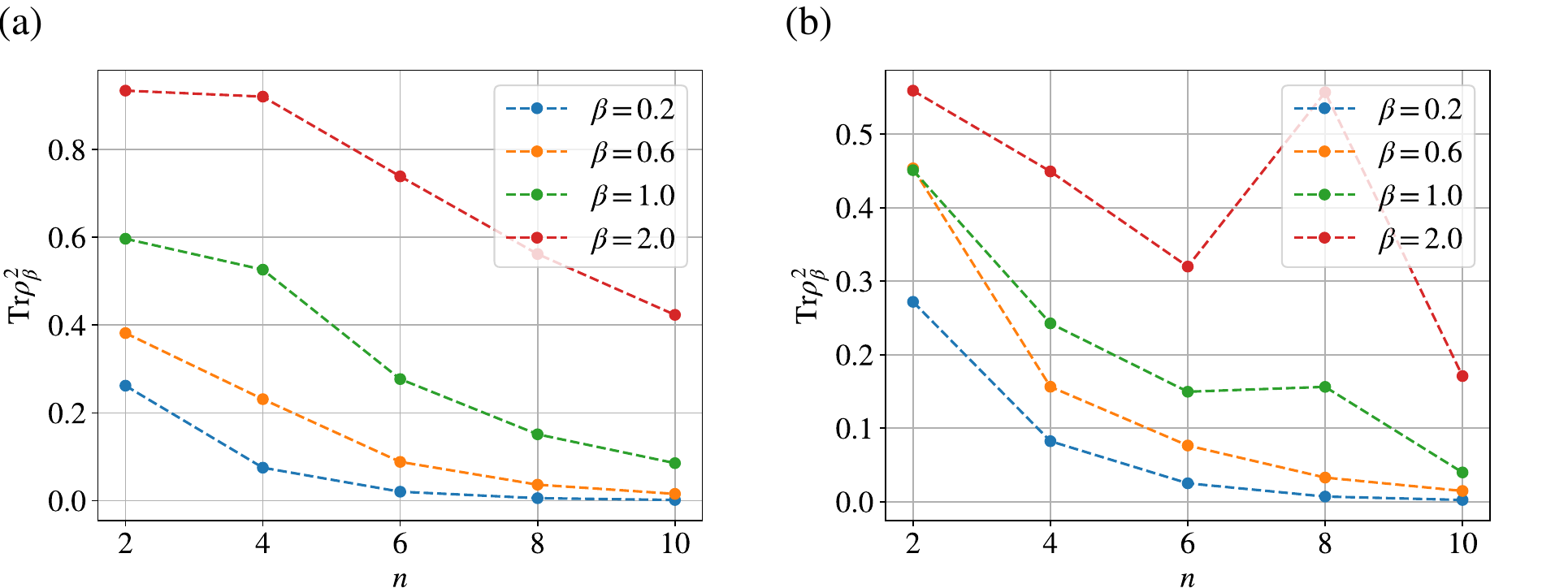}
    \caption{The purity, $\Tr{\rho_\beta^2}$, as a function of the system size, $n$, for (a) the XXZ-Heisenberg model (with $J=0.5$, $\Delta=0.7$, and closed boundary conditions) and (b) a random Hamiltonian with all-to-all connectivity (the XYZ-Heisenberg model with random couplings between all qubits and a random field on each qubit). Note that for (b), we use a different random initialization for each $n$. The jump observed at $n=8$ and $\beta=2$ is an outlier.}
    \label{fig:purity}
\end{figure}

First, in Fig.~\ref{fig:purity}, we show the purity, $\Tr{\rho_\beta^2}$, of the Gibbs states for two different types of Hamiltonians as a function of the number of qubits, $n$. We observe that for both Hamiltonians, the purity decreases (exponentially) with system size. The rate of decrease depends on the inverse temperature $\beta$. This is in accordance with our claim in the main text that, in general, the purity of the Gibbs state at finite temperature decreases with system size.

\begin{figure}[ht]\centering
    \includegraphics[width=0.46\columnwidth]{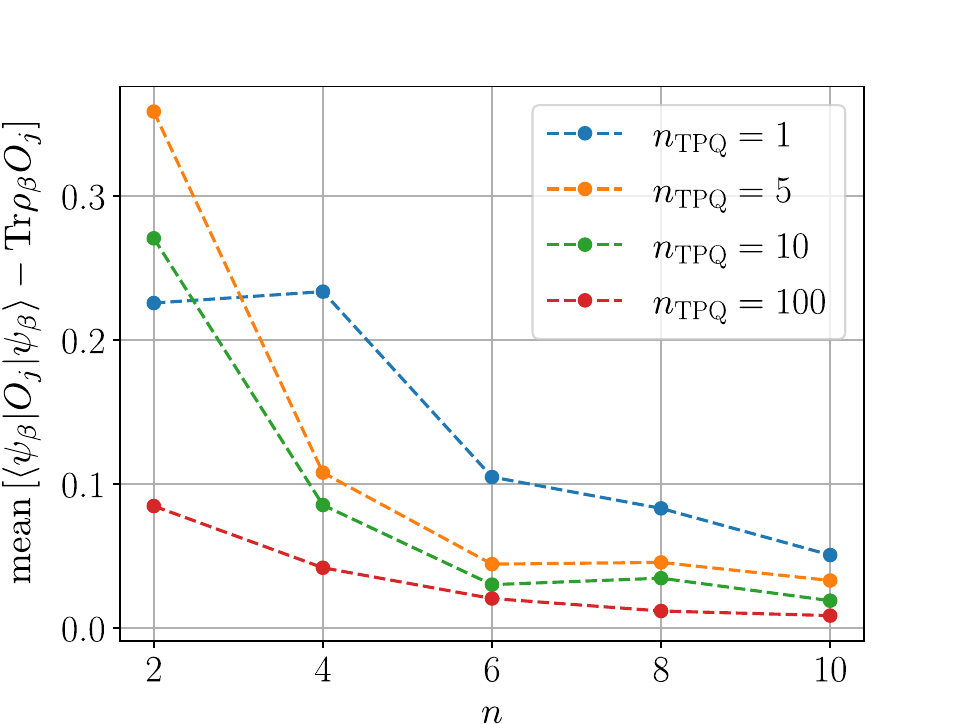}
    \caption{The mean error of an ensemble of size $n_{\text{TPQ}} \in \{1, 5,10, 100\}$ thermal pure states and different system sizes. For this, we compute the mean error on all possible one- and two-qubit Pauli observables. We see that for larger $n$, the required size of the ensemble, for a fixed mean error, becomes smaller, in accordance with the vanishing of the variance. For these simulations, we use the XXZ-Heisenberg model with $J=0.5$, $\Delta=0.7$, and closed boundary conditions.}
    \label{fig:tpq}
\end{figure}

In Fig.~\ref{fig:tpq}, we show the mean error of the expectation values of all possible one-qubit Pauli operators and all two-qubit Pauli XX, YY and ZZ operators, $\{O_j\}$, in a batch of TPQ states, $\mathbb{E}_U\left[\bra{\psi_\beta}O_j\ket{\psi_\beta}\right]$, as a function of the system size. Note that here the average, $\mathbb{E}_U$, is taken over a total of $n_{\mathrm{TPQ}}$ TPQ states. When we increase the system size, the mean error on all observables decreases for each different batch size, $n_{\mathrm{TPQ}}$. This means that for large system sizes, $n\geq 10$, it might be possible to get away with only one single TPQ as the estimator for the Gibbs state.

\subsection{Training of quantum Boltzmann machines}

In order to train a QBM, we minimize the quantum relative entropy, 
\begin{equation}\label{eq:qre}
    S(\eta \| \rho_\theta) = \text{Tr}(\eta \log{\eta}) - \text{Tr}(\eta \log{\rho_\theta}),
\end{equation}
between the QBM, $\rho_\theta=e^{-H(\theta)}/Z$, with $H(\theta)=\sum_{ij}\theta_{ij}H_{ij}$, and the target density matrix, $\eta$. The quantum relative entropy is exactly zero when $\rho_\theta=\eta$ and $S>0$ otherwise. For the minimization, we use vanilla gradient descent on the model parameters, $\theta \mapsto \theta - \alpha \nabla_\theta S(\eta \| \rho_\theta)$, with learning rate $\alpha > 0$. For this, we need the derivatives of Eq.~\eqref{eq:qre} with respect to the model parameters, which can be found to be~\cite{Amin_2018, Benedetti2017, Kieferova2016, Kappen2020}
\begin{equation}\label{eq:qbmgrad}
    \frac{\partial S(\eta \| \rho_\theta)}{\partial \theta_{ij}} = \langle H_{ij}\rangle_\eta -\langle H_{ij}\rangle_{\rho_\theta}.
\end{equation}
In practice, the first term in this derivative is usually given (experimental measurement data) or can be computed efficiently classically (for density matrices based on a classical data set, as discussed in Ref.~\cite{Kappen2020}). The second term, $\langle H_{ij}\rangle_{\rho_\theta}$, requires the evaluation of expectation values of the parametrized Gibbs state $\rho_\theta$. This is where we apply the pure thermal-shadow algorithm.
For the specific fully connected QBM with $H(\theta) = \sum_{k=x,y,z}\sum_{i,j>i}^{n}\lambda^k_{ij}\sigma_i^k\sigma^k_j + \sum_i^n \gamma^k_i\sigma_i^k$ discussed in the main text, we have 
\begin{equation}
     \frac{\partial S(\eta \| \rho_\theta)}{\partial \lambda_{ij}^k} = \langle \sigma_i^k\sigma^k_j\rangle_\eta -\langle \sigma_i^k\sigma^k_j\rangle_{\rho_\theta}, \hspace{2cm} \frac{\partial S(\eta \| \rho_\theta)}{\partial \gamma_{i}^k} = \langle \sigma_i^k\rangle_\eta -\langle \sigma_i^k\rangle_{\rho_\theta}.
\end{equation} 

\twocolumngrid

\bibliography{refs}

\end{document}